\documentstyle[12pt]{article}
\begin{document}

\title{Burgers velocity fields and the electromagnetic forcing
in  Schr\"{o}dinger's  interpolating dynamics}
 
\author{Piotr Garbaczewski, Grzegorz Kondrat, Robert Olkiewicz\\ 
Institute of Theoretical Physics, University of Wroc{\l}aw,
pl. M. Borna 9, \\
PL-50 204 Wroc{\l}aw, Poland}

\maketitle
\hspace*{1cm}

\begin{abstract}
We explore a connection of the deterministically  forced Burgers
equation for local velocity fields with  probabilistic
solutions of the  Schr\"{o}dinger
boundary data problem. An issue   of deducing the most likely
interpolating dynamics from the given initial and terminal
probability density data is here investigated  to give account
of  the perturbation by external  electromagnetic fields.
A suitable extension  of the Hopf-Cole logarithmic transformation
allows to deal with nonconservative drift fields and implies  the
validity of  appropriately  generalised heat equations, which
completely determine the  dynamics.

\end{abstract}

\newpage

\section{Burgers velocity fields and related  transport
processes}

The Burgers equation with (usually without, \cite{burg,hopf})
the forcing  term $\vec{F}(\vec{x},t)$:
$${\partial _t\vec{v} + (\vec{v}\nabla )\vec{v} = \nu 
\triangle \vec{v} +
\vec{F}(\vec{x},t)}\eqno (1)$$
and especially its statistically relevant $curl\, \vec{v}=0$
solutions, recently  have acquired  a considerable 
popularity in the variety of physical contexts.
 They  range from an  astrophysical issue
of the stratified large-scale distribution  of matter
in the early Universe, \cite{zeld,alb,woy}, through accoustic
turbulence dealing with 
intense noise in compressible liquids and gases, \cite{gurb}, to
primitive fluid turbulence modeling in terms of  the  statistics of 
Burgers shocks
in the low viscosity regime (enhanced by  random  initial data),
\cite{zeld,burg,frisch,sinai}, eventually  ending with the
analysis of a fully developed
Burgers turbulence that is  regarded as a result of random forcing
(stirring)  of respective velocity fields, \cite{fournier,woy1,parisi}.
It also pertains  to the turbulence-without-pressure  models,
\cite{pol},
description of directed polymers in a random potential, \cite{parisi},
 random interface growth problem, \cite{parisi1}, and 
fluctuations/dispersion   in deterministic   or random flows, 
\cite{parisi1,siggia,yak,monin}.
Last but not least, all  that implies  quite
nontrivial mathematics (very prohibitive with respect to the widely  
spread formal white noise manipulations), \cite{truman,holden,walsh}.

Although Burgers velocity fields  can be 
 analysed on their own, frequently one needs a supplementary 
insight into the matter transport dynamics that is
consistent with the  chosen (Burgers) velocity field evolution.
Then, the passive scalar (tracer or contaminant)  
advection-in-a-flow  problem, \cite{siggia,parisi,monin} naturally 
appears through the  parabolic dynamics:
$${\partial _tT + (\vec{v}\nabla )T = \nu \triangle T}\eqno (2)$$
Here, the Burgers velocity field defines the
flow which amplifies  the "passive scalar dispersion" (in  
 analogy to the standard turbulent transport effects). This in turn 
 implies   the time-evolution of the concentration $T(\vec{x},t)$ 
of the contaminant in the driving medium (incompressible fluid,
$\nabla \vec{v}=0$, but infinitely compressible gas as well).
While looking for the stochastic implementation of the microscopic 
(molecular) dynamics, \cite{woy2,parisi,monin},  it is  assumed
that the  "diffusing scalar" (contaminant  in the lore of early 
statistical turbulence models)  obeys an It\^{o} equation:  
$${d\vec{X}(t) =\vec{v}(\vec{x},t)dt  + \sqrt{2\nu } d\vec{W}(t)}
\eqno (3)$$
$$\vec{X}(0)=\vec{x}_0 \rightarrow \vec{X}(t)=\vec{x}$$
where the given  forced Burgers velocity field is    
perturbed by the noise term representing a molecular diffusion.
In the It\^{o} representation of diffusion-type random variable
$\vec{X}(t)$ one explicitly refers to the Wiener process  
$\sqrt {2\nu }\vec{W}(t)$, 
instead of the usually adopted formal white noise integral 
$\int_0^t \vec{\eta }(s)ds$, coming from the  Langevin-type version
of (3).

Under these premises, we cannot view equations (1)-(3) as 
completely independent (disjoint) problems:
the velocity field $\vec{v}$ cannot be  arbitrarily
inferred from  (1) or any other velocity-defining equation without
verifying the \it consistency \rm conditions, which would allow to
associate (2) and (3)  with 
a well defined random dynamics (stochastic process), and 
Markovian diffusion in particular, \cite{fried,horst}.
In connection with the usage of Burgers velocity fields (with
or without external forcing) which in (3) clearly are
intended to replace the 
standard \it forward  drift \rm of the would-be-involved Markov
diffusion process,  we have not found in the literature any 
attempt to resolve apparent contradictions arising if (2) and/or (3) 
are  defined by means of (1).

Moreover, rather obvious hesitation could have been  observed 
in attempts to establish the  most appropriate  matter transport 
rule,  if (1)-(3)  are adopted. 
Depending  on the particular phenomenological departure point,
 one either adopts the standard continuity equation,
 \cite{zeld,alb}, that is certainly
valid to a high degree of  accuracy in the low viscosity limit 
$\nu \downarrow 0$  of (1)-(3),  but incorrect on
mathematical grounds \it if \rm  there is a
diffusion involved \it and \rm simultaneously a solution of (1) stands 
for the respective \it current \rm velocity of the flow: 
${\partial _t\rho (\vec{x},t)= - \nabla 
[\vec{v}(\vec{x},t)\rho (\vec{x},t)]\enspace . }
$ 
Alternatively, following  the white noise calculus tradition telling
that  the stochastic integral
$\vec{X}(t)=\int_{0}^{t} \vec{v}(\vec{X}(s),s)ds +
\int_{0}^{t} \vec{\eta }(s)ds$
necessarily implies the Fokker-Planck equation, one adopts,
\cite{woy2}:
${\partial _t\rho (\vec{x},t) = \nu \triangle \rho (\vec{x},t) - 
\nabla [\vec{v}(\vec{x},t)\rho (\vec{x},t)]}$
which is clearly problematic in view of the classic Mc Kean's
discussion
of the  propagation of  chaos for the Burgers equation,
\cite{kean,cald,osa} and the derivation of the stochastic 
"Burgers process" in this  context: 
"the fun begins in trying to describe this Burgers motion as the 
path of a tagged molecule in an infinite bath of like molecules", 
\cite{kean}. 

Moreover, an issue of the necessary \it correlation \rm (cf.
\cite{monin},
Chap.7.3, devoted to the turbulent transport and the related
dispersion of contaminants)  between the probabilistic
Fokker-Planck
dynamics of the diffusing tracer, and this of the  passive tracer 
(contaminant)  concentration (2),  usually is left aside in the  
literature. 

To put things on the solid ground, let us consider a Markovian
diffusion  process, which is characterised by the
transition probability density
(generally inhomogeneous in space and time law of random
displacements)
$p(\vec{y},s,\vec{x},t)\, ,\, 0\leq s<t\leq T$, and the probability 
density $\rho (\vec{x},t)$  of its random variable 
$\vec{X}(t)\, ,\,  0\leq t \leq T$. The process is completely 
determined  by these data. For clarity of discussion, we do not
impose  any spatial boundary  restrictions, nor fix any concrete 
limiting value of $T$ which, in principle, can be moved to infinity.

The conditions valid for any $\epsilon >0$: \\
(a) there holds
$lim_{t\downarrow s}{1\over
{t-s}}\int_{|\vec{y}-\vec{x}|>\epsilon }
p(\vec{y},s,\vec{x},t)d^3x=0$,\\
(b) there exists a (forward) drift
$\vec{b}(\vec{x},s)=lim_{t\downarrow s}{1\over {t-s}}\int_{|\vec{y}-
\vec{x}|
\leq \epsilon }(\vec{y}-\vec{x})p(\vec{x},s,\vec{y},t)d^3y$, \\
(c) there exists  a diffusion function (in our case  it is simply
a diffusion coefficient $\nu $)
$a(\vec{x},s)=lim_{t\downarrow s}{1\over {t-s}}
\int_{|\vec{y}-\vec{x}|\leq \epsilon }  (\vec{y}-\vec{x})^2
p(\vec{x},s,\vec{y},t)d^3y$,\\
are conventionally interpreted to define a diffusion process,
\cite{horst,fried}.
Under suitable restrictions (boundedness of involved
functions, their continuous differentiability) the function:
$${g(\vec{x},s)=E\{g(\vec{X}(T))|\vec{X}(s)=\vec{x}; s\leq T\}=
\int  p(\vec{x},s,\vec{y},T)g(\vec{y},T) d^3y }\eqno (4)$$
satisfies the backward diffusion equation (notice two sign 
changes in comparison  with (2))
$${- \partial _sg(\vec{x},s) = \nu \triangle g(\vec{x},s)  +
[\vec{b}(\vec{x},s)\nabla ]g(\vec{x},s)
\enspace .}\eqno (5)$$
Let us point out that the validity of (5) is known to be a \it
necessary
 \rm condition for the existence of a Markov diffusion process, whose
probability density $\rho (\vec{x},t)$ is to obey the Fokker-Planck 
equation (the forward drift $\vec{b}(\vec{x},t)$ replaces the
previously  utilized Burgers velocity  $\vec{v}(\vec{x},t)$):
$${\partial _t\rho (\vec{x},t) = \nu \triangle 
\rho (\vec{x},t) - \nabla [\vec{b}(\vec{x},t)\rho (\vec{x},t)]}
\eqno (6)$$

The case  of particular interest, in the traditional 
nonequilibrium statistical physics literature,  
appears when $p(\vec{y},s,\vec{x},t)$
is  a \it fundamental solution \rm of (5) with respect to variables
$\vec{y},s$,  \cite{krzyz,fried,horst},
see however \cite{olk2} for  an analysis of alternative situations.
Then, the transition probability density satisfies \it also \rm
the second Kolmogorov (e.g. the Fokker-Planck) equation in the
remaining $\vec{x}, t$ pair of variables.  Let us emphasize that
these two equations form an adjoint pair of partial differential
equations, referring to the  slightly counterintuitive for
physicists, though transparent for
mathematicians, \cite{haus,fol,has,nel,zambr,zambr1}, issue of time
reversal of diffusions.

After  adjusting (3) to the present  context, $\vec{X}(t)=
\int_0^t\, \vec{b}(\vec{X}(s),s)\, ds + \sqrt {2\nu } \vec{W}(t) $
we can utilize the standard rules of the It\^{o} stochastic calculus,
\cite{nel1,nel,zambr,zambr1},  to realise that for any smooth function
$f(\vec{x},t)$ of the random variable $\vec{X}(t)$ the 
conditional expectation value:
$$lim_{\triangle t\downarrow 0} {1\over {\triangle t}}\bigl [\int
p(\vec{x},t,\vec{y},t+
\triangle
t)f(\vec{y},t+\triangle t)d^3y - f(\vec{x},t)\bigr ] = 
(D_+f)(\vec{X}(t),t)=    \eqno (7) $$
$$=  (\partial _t + \vec{b}\nabla + \nu \triangle )f(\vec{x},t)
\enspace , $$
where $\vec{X}(t)=\vec{x}$,
determines  the forward drift $\vec{b}(\vec{x},t)$ (if we set 
components of $\vec{X}$ instead of $f$) and allows to introduce  
the local field of (forward) accelerations associated with the 
diffusion process, which we \it constrain \rm by demanding
(see e.g. Refs.
\cite{nel,nel1,zambr,zambr1} for prototypes of such dynamical 
constraints):
$${(D^2_+\vec{X})(t) =(D_+\vec{b})(\vec{X}(t),t) =(\partial _t
\vec{b} +
(\vec{b}\nabla )\vec{b} + \nu \triangle
\vec{b})(\vec{x},t)= \vec{F}(\vec{x},t)}\eqno (8)$$
where $\vec{X}(t)=\vec{x}$ and, at the moment arbitrary, 
function $\vec{F}(\vec{x},t)$ may be  interpreted as
an external forcing applied to  the diffusing system,
\cite{blanch}. 

In particular, if we assume that drifts remain
gradient fields, $curl \, \vec{b}= 0$, under the forcing, then
those  that are allowed by the prescribed choice of
$\vec{F}(\vec{x},t)$  \it must \rm fulfill the compatibility
condition  (notice the conspicuous absence of the standard
Newtonian minus sign in this analogue of the second Newton law,
see e.g. \cite{risken,zambr,blanch})
$${\vec{F}(\vec{x},t)=\nabla \Omega (\vec{x},t) }\eqno (9)$$
$$\Omega (\vec{x},t) = 2\nu \bigl [\partial _t \Phi \, +\,
{1\over 2} ({b^2
\over {2\nu }}+ \nabla b)\bigr ]$$
which establishes the Girsanov-type martingale connection of
the forward drift
$\vec{b}(\vec{x},t)=2\nu \nabla \Phi (\vec{x},t)$ with the
Feynman-Kac,
cf. \cite{blanch,olk2}, potential $\Omega (\vec{x},t)$ of the
chosen external force field.

One of the distinctive features of Markovian diffusion processes
with the positive density $\rho (\vec{x},t)$ is that the
notion of the \it backward \rm transition probability density
 $p_*(\vec{y},s,\vec{x},t)$ can be consistently introduced on 
each finite  time interval, say $0\leq s<t\leq T$:
 $${\rho (\vec{x},t) p_*(\vec{y},s,\vec{x},t)=p(\vec{y},s,\vec{x},t)
 \rho (\vec{y},s)} \eqno (10)$$
so that $\int \rho (\vec{y},s)p(\vec{y},s,\vec{x},t)d^3y=
\rho (\vec{x},t)$ 
and $\rho (\vec{y},s)=\int p_*(\vec{y},s,\vec{x},t)
\rho (\vec{x},t)d^3x$.  
This  allows to define the backward derivative of the process in the
conditional mean (cf. \cite{blanch,vig,olk} for a discussion of
these concepts in case of the most traditional Brownian motion and
Smoluchowski-type diffusion processes)
$$lim_{\triangle t\downarrow 0} \, {1\over {\triangle t}}\bigl
[ \vec{x} - \int p_* (\vec{y},t-\triangle t,\vec{x},t)\vec{y} d^3y
\bigr ]= (D_-\vec{X})(t)=
\vec{b}_*(\vec{X}(t),t)
\eqno (11) $$
$$(D_-f)(\vec{X}(t),t) = (\partial _t + \vec{b}_* \nabla - \nu
\triangle )f(\vec{X}(t),t)$$
Accordingly, the backward version  of the acceleration field reads
$${(D^2_-\vec{X})(t) = (D^2_+\vec{X})(t) = \vec{F}(\vec{X}(t),t) }
\eqno (12) $$
where  in view of $\vec{b}_*= \vec{b} - 2\nu \nabla ln \rho $
we have  explicitly fulfilled  the forced Burgers equation
(cf. (1)):
$${\partial _t\vec{b}_* +  (\vec{b}_*\nabla )\vec{b}_* - 
\nu \triangle \vec{b}_* 
= \vec{F}} \eqno (13)$$
and, \cite{nel,zambr,blanch},  under the gradient-drift field
assumption, $curl \, \vec{b}_*=0$, we deal
with $\vec{F}(\vec{x},t)$ given by (9) again.
A notable consequence of the involved   backward 
It\^{o} calculus  is that the Fokker-Planck equation (6) can be 
transformed to an \it equivalent \rm  form of:
$${\partial _t\rho(\vec{x},t) = - \nu \triangle \rho (\vec{x},t) - 
\nabla [\vec{b}_*(\vec{x},t) \rho (\vec{x},t)]}\eqno (14)$$
with the very same initial (Cauchy) data 
$\rho _0(\vec{x})=\rho (\vec{x},0)$  as in  (6).  

At this point let us recall that  equations (5) and (6) form a natural 
 adjoint  pair of equations for the  diffusion process  
 in the time interval $[0,T]$.  
Clearly, an adjoint of  (14) reads:
$${\partial _s f(\vec{x},s) = \nu \triangle f(\vec{x},s) - 
[\vec{b}_*(\vec{x},s)\nabla ] f(\vec{x},s)}\eqno (15)$$
where:
$${f(\vec{x},s)=\int p_*(\vec{y},0,\vec{x},s) f(\vec{y},0)d^3y
\enspace  , }\eqno (16)$$
to be compared with (4),(5) and the previously mentioned 
passive scalar dynamics (2), see e.g. also \cite{woy2}. 
Here, manifestly, the time evolution of 
the backward drift is governed by the Burgers equation, 
and the diffusion equation (15) is correlated (via the 
definition (10)) with the probability density evolution rule (14). 

This pair \it only \rm can be consistently utilized if the 
diffusion proces  is to be  driven by forced (or unforced) 
Burgers velocity fields. 

On the other hand, the study of diffusion driven by  
the Burgers flow may begin from first solving the Burgers equation
(12) for a chosen external force field, next 
specifying the  probability density (14), 
eventually ending with the corresponding "passive 
contaminant" concentration dynamics (15), (16). 
All that  remains in a perfect 
 agreement with  the heuristic discussion of the 
concentration dynamics given in   Ref. \cite{monin}, Chap. 7.3. 
where the "backward dispersion" problem 
with "time running backwards" was found necessary to \it predict \rm 
the concentration.

Let us notice that the familiar logarithmic Hopf-Cole transformation, 
\cite{hopf,flem}, of the Burgers equation into the 
generalised diffusion equation (yielding explicit solutions in the 
unforced case) has received a generalisation in the framework of the 
so called Schr\"{o}dinger interpolation problem, 
\cite{zambr,zambr1,olk2,blanch,olk,olk1}, see also \cite{alb1,freid}.
In its recent reformulation, \cite{olk2}, the solution in terms 
of the interpolating  Markovian diffusion process is found to rely 
on the  adjoint pairs  of parabolic equations, like e.g. (5), (6) 
or (14), (15). 
In case of gradient drift fields the process can be determined
by   checking  (this imposes limitations on the admissible 
force field potential) whether  the Feynman-Kac kernel 
$${k(\vec{y},s,\vec{x},t)=\int exp[-\int_s^tc(
\vec{\omega }(\tau ),\tau)d\tau ]
d\mu ^{(\vec{y},s)}_{(\vec{x},t)}(\omega )}\eqno (17) $$
is positive and continuous in the open space-time
area of interest (then, additional limitations on the path measure
need to be introduced, \cite{blanch,chung}), and whether it
gives rise to
positive solutions of the adjoint pair of  generalised 
heat equations:
$${\partial _tu(\vec{x},t)=\nu \triangle u(\vec{x},t) -
c(\vec{x},t)u(\vec{x},t)}\eqno (18)$$
$$\partial _tv(\vec{x},t)= -\nu \triangle v(\vec{x},t) +
c(\vec{x},t)v(\vec{x},t)$$
Here, a function $c(\vec{x},t)={1\over {2\nu }} \Omega (\vec{x},t)$
is restricted only by the positivity and continuity demand
for the kernel (17),  see e.g.  \cite{freid,olk2}.
In the above, $d\mu ^{(\vec{y},s)}_{(\vec{x},t)}(\omega)$ is the
conditional
Wiener measure over sample paths of the standard Brownian motion.

Solutions of (18), upon suitable normalisation give rise to the
Markovian  diffusion process with the factorised probability density
$\rho (\vec{x},t)=u(\vec{x},t)v(\vec{x},t)$ wich interpolates between 
the boundary density data $\rho (\vec{x},0)$ and 
$\rho (\vec{x},T)$, with the forward and backward 
drifts of the process defined as follows:
$${\vec{b}(\vec{x},t)=2\nu {{\nabla v(\vec{x},t)}
\over {v(\vec{x},t)}}}
\eqno (19)$$
$$\vec{b}_*(\vec{x},t)= - 2\nu {{\nabla u(\vec{x},t)}
\over {u(\vec{x},t)}}$$
in the prescribed time interval $[0,T]$.  These formulas imply that
the compatibility condition  (9),  connecting the drifts with the
a priori chosen function  $c(\vec{x},t)=
{1\over {2\nu }}\Omega (\vec{x},t)$, can be easily established.

The transition probability density of this process reads:
$${p(\vec{y},s,\vec{x},t)=k(\vec{y},s,\vec{x},t)
{{v(\vec{x},s)}\over {v(\vec{y},t)}}\enspace .}\eqno (20)$$
while the corresponding (since $\rho (\vec{x},t)$ is given) transition  
probability density, (10),  of the backward process has the form:
$${p_*(\vec{y},s,\vec{x},t) = k(\vec{y},s,\vec{x},t)
{{u(\vec{y},s)}\over {u(\vec{x},t)}}\enspace .}\eqno (21)$$
Obviously, \cite{olk2,zambr}, in the time interval $0\leq s<t\leq T$
there holds:
$${u(\vec{x},t)=\int u_0(\vec{y}) k(\vec{y},s,\vec{x},t) d^3y}
\eqno (22)$$
$$v(\vec{y},s)=\int k(\vec{y},s,\vec{x},T) v_T(\vec{x})d^3x \enspace .
$$

By defining $\Phi _*=log\, u$, 
we immediately recover the traditional form of the Hopf-Cole
transformation 
 for Burgers velocity fields: $\vec{b}_*=-2\nu \nabla \Phi _*$.
In the special case of the standard free Brownian motion, there 
holds $\vec{b}(\vec{x},t)=0$ while $\vec{b}_*(\vec{x},t)=-2\nu 
\nabla log\, \rho (\vec{x},t)$.

\section{The problem of electromagnetic forcing in the
Schr\"{o}\-din\-ger interpolating dynamics}

It turns out the the crucial point of our previous discussion 
lies in a \it proper \rm choice of the strictly positive and
continuous, in an open space-time area,  
function $k(\vec{y},s,\vec{x},t)$ which, if we wish to 
construct a Markov  process, 
has to satisfy the Chapman-Kolmogorov (semigroup composition) 
equation.  It has led us to 
consider a pair of adjoint partial differential equations, (18),
as an alternative to either (5), (6) or (14), (15).  

In the quantally oriented literature dealing with Schr\"{o}dinger
operators and their spectral properties, \cite{simon,glimm,chung},
the potential $c(x,t)$ in (17), (18)
is usually assumed  to be a  continuous and bounded from below
function, but these restrictions can be substantially relaxed
(unbounded functions are allowed in principle) if we wish to
consider  general Markovian diffusion processes  and disregard
an issue of  the bound state spectrum and this of the ground state
 of the  (selfadjoint) semigroup generator, \cite{krzyz,fried}.
Actually, what we need is merely that properties of $c(\vec{x},t)$
allow for the kernel $k$, (17), which  is positive and continuous.
Taking for granted that suitable
conditions are fulfilled, \cite{simon,olk2},
we can immediately associate with  equations (18) an
integral kernel of the time-dependent semigroup
(the exponential operator should be understood as  time-ordered
expression, since in general $H(\tau )$ may not commute with
$H(\tau ')$ for $\tau \neq \tau '$):
$${k(\vec{y},s,\vec{x},t)=
[exp(-\int_s^t H(\tau )d\tau )](\vec{y},\vec{x})}\eqno (23)$$
where $H(\tau )=-\nu \triangle +c(\tau )$ is the pertinent
semigroup generator.
Then, by the Feynman-Kac formula, \cite{freid}, we get an
expression (17) for the kernel, which in turn yields (19)-(22),
see e.g. \cite{olk2}.
As mentioned before, (20) combined with (17) sets a Girsanov
probabilistic (martingale) connection between the Wiener measure
corresponding to the standard Brownian motion with
$\vec{b}(\vec{x},t)=0$, and this appropriate
for the diffusion process with a nonvanishing  drift
$\vec b(\vec{x},t),\, curl \, \vec{b} =0$. 

The above formalism is known, \cite{blanch},
to encompass the standard
Smoluchowski-type diffusions in conservative force fields.

Strikingly, an investigation of electromagnetically forced diffusions 
has not been much pursued in the literature, although an issue of 
deriving the Smoluchowski-Kramers equation (and possibly its large 
friction limit) from the Langevin-type equation for the charged 
Brownian particle in the general electromagnetic field  has 
been relegated in Ref. \cite{schuss}, Chap. 6.1 to the status of 
the innocent-looking exercise (sic !).  
On the other hand, the diffusion
of realistic charges in dilute ionic solutions creates a number of 
additional difficulties due to the apparent Hall mobility in terms 
of mean currents induced by the electric field (once assumed to act 
upon the system), see e.g. \cite{hub,sung} and \cite{mori}.

In connection with the electromagnetic forcing of diffusing charges, 
the gradient field
assumption imposes  a severe limitation if we account for typical
(nonzero circulation) features of the classical
motion  due to the Lorentz force, with or  without the random
perturbation component.  
The purely electric forcing is simpler to handle, since it has 
a definite gradient field  realisation, see e.g. \cite{izm} for a
recent discussion of related issues.
The major obstacle with respect to our previous (Section 1)
discussion is that, if we wish to regard either
the force  $\vec{F}$, (8), (12),
or  drifts $\vec{b}$, $\vec{b}_*$  to have an electromagnetic
origin, then necessarily we need to pass from conservative
to non-conservative  fields. This subject matter has not been
significantly exploited so far in the nonequilibrium statistical
physics literature.

Interestingly, in the framework of the Onsager-Machlup approach
towards an identification of most probable paths  which one can
associate with the underlying diffusion process,
\cite{has1,hunt,ronc}, the non-conservative model system has
been investigated, \cite{wieg}.
Namely, an  effectively  two-dimensional Brownian motion was
analyzed, whose three-dimensional forward
drift $\vec{b}(\vec{x}), \, b_3=0$ in view of
$\partial_xb_1\neq \partial _yb_2$
has $curl \, \vec{b}\neq 0$.
Then, by the standard variational
argument with respect to the Wiener-Onsager-Machlup action,
\cite{hunt,wieg},
$${I\{ L(\dot{\vec{x}},\vec{x},t);t_1,t_2\} = {1\over {2\nu }}
\int_{t_1}^{t_2} \{ {1\over 2}[\dot{\vec{x}} -
\vec{b}(\vec{x},t)]^2 +
\nu \nabla \vec{b}(\vec{x},t)\} dt \enspace ,}\eqno (24)$$
 the most probable trajectory, about which  major contributions
from  (weighted)  Brownian paths are concentrated,
was found to be a solution of the Euler-Lagrange equations,
which are formally identical to the equations of motion
$${\ddot{\vec{q}}_{cl}= \vec{E} +
\dot{\vec{q}}_{cl} \times \vec{B}}\eqno (25)$$
of a classical particle of unit mass and
unit charge moving in an electric field $\vec{E}$ and the magnetic
field $\vec{B}$.
The electric field (to be compared with Eq. (9)) is given by:
$${\vec{E}=-\nabla \Phi }\eqno (26)$$
$$\Phi = -{1\over 2} (\vec{b}^2 + 2\nu \nabla \vec{b})$$
while the magnetic field has the only nonvanishing component in
 the z-direction of $R^3$:
$${\vec{B}=curl \, \vec{b}= \{ 0,0,\partial _xb_2-\partial _yb_1\}
\enspace , } \eqno (27)$$
 Clearly, $\vec{B}= curl \vec{A}$ where
$\vec{A}\dot{=}\vec{b}$ is the electromagnetic vector potential.
The simplest  example  is a notorious constant magnetic
field defined by  $b_1(\vec{x})=-{B\over 2} x_2,\, b_2(\vec{x})=
{B\over 2} x_1$.

One immediately realizes that the Fokker-Planck equation in
this case is incompatible with  traditional intuitions
underlying the Smoluchowski-drift identification: the forward
drift is not proportional to  an external force, but to an
electromagnetic potential.
Nevertheless, the variational
information  drawn from the Onsager-Machlup Lagrangian involves
the Lorentz force-driven trajectory.
Hence, some principal effects of the
electromagnetic forcing are present in the diffusing system, whose
drifts display an "unphysical" (gauge dependent) form.

On the other hand,  if we accept this "unphysical" random motion
to yield the representation
$d\vec{X}(t)=\vec{A}(\vec{X}(t),t)dt +\sqrt{2\nu } d\vec{W}(t)$,
supplemented by
the corresponding pair (5), (6) of adjoint diffusion equations
with $\vec{A}(\vec{x},t)$  replacing $\vec{b}(\vec{x},t)$,
then standard rules of the It\^{o} stochastic calculus, (8),
tell us that
$${(D^2_+\vec{X})(t)=\partial _t\vec{A} + (\vec{A}\nabla )\vec{A} +
\nu \triangle \vec{A} = - {{B^2}\over 2} \{ x_1,x_2,0 \} = -
\vec{E}(\vec{x})}\eqno (28)$$
where $\vec{E}(\vec{x})={{B^2}\over 2}\{ x_1,x_2,0 \} $, if
calculated from (26).

We thus arrive at the purely electric forcing with
reversed sign (if compared with that coming from the
Onsager-Machlup argument, (26)) and somewhat surprisingly,
there is no   impact of the previously discussed magnetic motion
 on the level of dynamical  constraints  (8), (13).
The adopted recipe is thus incapable of producing the magnetically
forced  diffusion process  that conforms with arguments of
Section 1.

In below, we shall follow more abstract route,
based on invoking the Feynman-Kac kernel idea (23), \cite{olk2}.
This approach has a clear  advantage of elucidating the
generic issues that hamper attempts to describe
electromagnetically forced diffusion processes.

Usually,  the selfadjoint semigroup generators attract the
attention of physicists in connection with the Feynman-Kac
formula. A typical route towards
incorporating electromagnetism comes from quantal motivations
via the minimal electromagnetic coupling recipe which
preserves the selfadjointness of the  generator (Hamiltonian
of the system).  As such, it constitutes a part of the general
theory of Schr\"{o}dinger operators.
A rigorous study of operators of the form
$-\triangle +V$ has become a well developed mathematical
discipline, \cite{simon}.
The study of Schr\"{o}dinger operators with magnetic fields,
typically of the form $-(\nabla -i\vec{A})^2 +V$,  is
less advanced, although specialised chapters on the magnetic field
issue can be found in monographs devoted to  functional
integration methods, \cite{simon,roep}, mostly in  reference  to
 seminal papers \cite{simon1,simon2}.

From the mathematical  point of view, it is desirable to deal with
 magnetic fields that go to zero at infinity, which is
certainly acceptable on physical grounds as well.  The
constant magnetic field (see e.g. our previous considerations)
does not meet this requirement, and its notorious usage in the
literature makes us (at the moment) to decline the asymptotic
assumption and inevitably fall into a number of serious
complications.

One obvious  obstacle can be seen immediately by taking
advantage of the
existing results, \cite{simon1}.  Namely, an explicit expression
for  the Feynman-Kac kernel in a constant magnetic  field,
introduced through the  the minimal electromagnetic coupling
assumption $H(\vec{A})=-{1\over 2}(\nabla -i\vec{A})^2$, is
available (up to irrelevant dimensional constants):
$${exp[- t H(\vec{A})](\vec{x},\vec{y})=
[{B\over {4\pi sinh({1\over 2}Bt)}}] ({1\over {2\pi t}})^{1/2}}
\eqno (29)$$
$$\times exp\{ - {1\over {2t}}(x_3-y_3)^2 - {B\over 4}
coth({B\over 2}t) [(x_2-y_2)^2+ (x_1-y_1)^2] -
i {B\over 2}(x_1y_2-x_2y_1)\} \enspace. $$
Clearly, it is  \it not  \rm   real (hence \it non-positive
\rm  and directly at variance with  the  major demand in the
Schr\"{o}dinger interpolation problem, as outlined in Section 1),
except for directions $\vec{y}$ that  are parallel to  a chosen
$\vec{x}$.

Consequently, a bulk of the well developed mathematical theory is
of no use for our purposes  and new techniques must be developed
for  a consistent description  of the electromagnetically
forced diffusion processes along the lines of Section 1, i.e.
within the framework of Schr\"{o}dinger's interpolation problem.

\section{Forcing via Feynman-Kac semigroups}

The conditional Wiener measure
$d\mu ^{(\vec{y},s)}_{(\vec{x},t)}(\vec{\omega })$,
appearing  in the Feynman-Kac kernel definition (17), if unweighted
(set $c(\vec{\omega }(\tau ),\tau)=0$)   gives rise to the
familiar heat kernel. This, in turn, induces the Wiener measure
$P_W$ of the set of
all sample paths,which originate from $\vec{y}$ at time $s$ and
terminate (can
be located) in the Borel set $A \in R^3$ after time $t-s$:
$P_W[A] = \int_A d^3x \int
d\mu ^{(\vec{y},s)}_{(\vec{x},t)}(\vec{\omega })
=\int_A d\mu $
where, for simplicity of notation, the $(\vec{y},t-s)$ labels are
omitted and $\mu ^{(\vec{y},s)}_{(\vec{x},t)} $ stands
for the heat kernel.

Having defined an It\^{o} diffusion
$\vec{X}(t)=\int_0^t \vec{b}(\vec{x},u)du + \sqrt{2\nu }
\vec{W}(t)$
we are interested in the analogous
path measure:
$P_{\vec{X}}[A] =
\int_A dx\int d\mu _{(\vec{x},t)}^{(\vec{y},s)}(
\vec{\omega }_{\vec{X}}) = \int _A d\mu (\vec{X})$.

Under suitable (stochastic, \cite{blanch}) integrability
conditions imposed on the forward drift, we have granted
the  absolute continuity $P_X \ll P_W$ of measures,
which implies the existence of a
strictly positive Radon-Nikodym density.  Its  canonical
Cameron-Martin-Girsanov form, \cite{simon,blanch}, reads:
$${{{d\mu (\vec{X})}\over {d\mu }}(\vec{y},s,\vec{x},t) =
exp  {1\over {2\nu }}\bigl [\int_s^t
\vec{b}(\vec{X}(u),u)d\vec{X}(u) - {1\over 2}
\int_s^t [\vec{b}(\vec{X}(u),u)]^2
du\bigr ] \enspace . } \eqno (30) $$

If we assume that drifts are gradient fields, $curl \, \vec{b}=0$,
then  the It\^{o} formula allows to reduce,
otherwise troublesome, stochastic
integration in the exponent of (30), \cite{simon,roep},  to
ordinary Lebesgue integrals:
$${{1\over {2\nu }}\int_s^t \vec{b}(\vec{X}(u),u)d\vec{X}(u)
= \Phi  (\vec{X}(t),t) - \Phi (\vec{X}(s),s) -
\int_s^t du\, [\partial _t\Phi  + {1\over 2} \nabla \vec{b} ]
(\vec{X}(u),u) \enspace .  } \eqno (31) $$
After inserting (31) to (30) and next integrating with respect to
 the conditional Wiener measure, on account of (9) we
 arrive at the standard   form of the Feynman-Kac kernel (17).
Notice that (31) establishes a probabilistic basis for logarithmic
transformations (19) of forward and backward drifts:
$b=2\nu \nabla log\, v=2\nu \nabla \Phi $,
$b_*=-2\nu \nabla log \, u=-2\nu \nabla \Phi _*$. The forward
version is commonly  used in connection with the
transformation of the Fokker-Planck equation  into the
generalised heat equation, \cite{risken,garb,blanch}.
The backward
version  is the Hopf-Cole transformation, mentioned in
Section 1, used to map the Burgers equation into the very
same generalised heat equation as in the previous case,
\cite{hopf,alb1}.

 However, presently we are interested in non-conservative
 drift fields, $curl \, \vec{b} \neq 0$, and in that case the
 stochastic integral in (30) is the major source of computational
 difficulties, \cite{simon,roep,nel}, for nontrivial vector
 potential field configurations.  It explains the virtual
 absence of magnetically forced diffusion problems in the
 nonequilibrium statistical physics literature.

 At this point, some  steps of the analysis performed in
 Ref. \cite{abc} in the context of the "Euclidean quantum
 mechanics", cf. also \cite{zambr1}, are extremely useful.
 Let us emphasize that  electromagnetic fields we utilize,  are
 always meant to be ordinary Maxwell fields with \it no \rm
 Euclidean connotations
 (see e.g. Chap.9 of Ref. \cite{roep} for the Euclidean version of
 Maxwell theory).

Let us consider a gradient drift-field diffusion problem  according
to Section 1, with (17), (31)  involved  and thus an  adjoint pair
(18) of parabolic equations completely defining the Markovian
diffusion process. Furthermore, let $\vec{A}(\vec{x})$ be the
time-independent vector potential for the Maxwellian
magnetic field $\vec{B}= curl  \, \vec{A}$.
We pass from the gradient realisation of drifts to  the
new  one, generalizing (19), for  which  the following
decomposition
into the gradient and   nonconservative part is valid:
$${\vec{b}(\vec{x},t)= 2\nu \, log\, \Phi (\vec{x},t) -
\vec{A}(\vec{x})\enspace ,} \eqno (32)$$
We  denote  $\theta (\vec{x},t)\dot{=}
exp\, [\Phi (\vec{x},t)]$
and admit that (32)  is a forward drift of an It\^{o} diffusion
process with a stochastic differential
${ d\vec{X}(t)= [2\nu {{\nabla \theta }\over {\theta }} -
\vec{A}]dt +  \sqrt {2\nu } d\vec{W}(t)}$.
On purely formal grounds, we deal here with  an example of the
Cameron-Martin-Girsanov transformation of the forward drift of
a given Markovian diffusion process and we are entitled to ask for
a corresponding measure transformation, (30).

To this end, let us furthermore  \it assume \rm that
$\theta (\vec{x},t)=\theta $ solves a partial differential equation
$${\partial _t\theta  =
 - \nu [ \nabla   - {1\over {2\nu }}
\vec{A}(\vec{x})]^2\theta  + c(\vec{x},t)\theta }\eqno  (33) $$
with the notation $c(\vec{x},t)={1\over {2\nu }}\Omega (\vec{x},t)$
patterned after (9).
 Then, by using  the It\^{o} calculus
and (32), (33) on the way, see e.g. Ref. \cite{abc}, we can
rewrite (30) as follows:
$${{{d\mu (\vec{X})}\over {d\mu }}(\vec{y},s,\vec{x},t)=}\eqno (34)$$
$${= exp\, {1\over {2\nu }} \bigl [
\int_s^t\bigl ( 2\nu {{\nabla \theta }\over {\theta }} -
\vec{A} \bigr ) (\vec{X}(u),u) d\vec{X}(u) - {1\over 2}
\int_s^t \bigl ( 2\nu {{\nabla \theta }\over {\theta }} -
\vec{A} \bigr )^2(\vec{X}(u),u)\, du\bigr ] }$$
$${=\, {{\theta (\vec{X}(t),t)}\over {\theta (\vec{X}(s),s)}}
\, exp\bigl [ - {1\over {2\nu }}  \int_s^t  [
\vec{A}(u)d\vec{X}(u) \, + \, \nu (\nabla \vec{A})(\vec{X}(u))du \,
+\, \Omega (\vec{X}(u),u)du ] \bigr ]\enspace , } $$
where $\vec{X}(s)=\vec{y}, \vec{X}(t)=\vec{x}$.

More significant observation is that  the Radon-Nikodym density (34),
if integrated  with respect to the conditional Wiener  measure,
gives rise to the Feynman-Kac kernel (23) of the \it
non-selfadjoint \rm
semigroup  (suitable integrability conditions need
to be respected here as well, \cite{abc}), with the
generator $H_{\vec{A}}=
 - \nu  [\nabla  - {1\over {2\nu }}
\vec{A}(\vec{x})]^2 + c(\vec{x},t)$
defined by the right-hand-side of (33):
$${\partial _t\theta (\vec{x},t) = H_{\vec{A}}\theta (\vec{x},t)=}$$
$${
= [- \nu \triangle + \vec{A}(\vec{x}) \nabla + {1\over 2} (\nabla
\vec{A}(\vec{x})) -
{1\over {4\nu }}[\vec{A}(\vec{x})]^2 +
c(\vec{x},t)]\theta (\vec{x},t)}\eqno (35)$$
$$= -\nu \triangle \theta (\vec{x},t)  +
\vec{A}(\vec{x})\nabla \theta (\vec{x},t) +
c_{\vec{A}}(\vec{x},t)\theta (\vec{x},t)
\enspace .$$
Here:
$${c_{A}(\vec{x},t)= c(\vec{x},t) +  {1\over 2}(\nabla
\vec{A})(\vec{x}) - {1\over {4\nu }} [\vec{A}(\vec{x})]^2\enspace .}
\eqno (36)$$
An adjoint parabolic partner of (35) reads:
$${\partial _t\theta _* = - H^*_{\vec{A}}
\theta _*  =
\nu \triangle  \theta _* +
\nabla [\vec{A}(\vec{x})\theta _*] -
c_{A}(\vec{x},t)\theta _* = }\eqno (37)  $$
$$= \nu [\nabla + {1\over {2\nu }}\vec{A}(\vec{x})]^2
\theta _*
- c(\vec{x},t)\theta _*\enspace .$$

Consequently, our assumptions (32), (33) involve  a
generalization of the adjoint parabolic system (18) to a new
adjoint one comprising  (33), (37). Obviously, the original form
of (18) is immediately restored by setting
$\vec{A}=\vec{0}$, and executing obvious replacements
$\theta _* \rightarrow u$, $\theta \rightarrow v$.

Let us emphasize again, that in contrast to Ref. \cite{abc}, where
the non-Hermitean generator $2\nu H_{\vec{A}}$, (33), has been
introduced as "the Euclidean version of the Hamiltonian"
$H=-2\nu ^2(\nabla -{i\over {2\nu }}\vec{A})^2 + \Omega $, our
electromagnetic fields stand for solutions of the usual Maxwell
equations and \it are not \rm Euclidean at all.

As long as the coefficient functions
(both additive and multiplicative) of the adjoint
parabolic system (35), (37) are not  specified, we remain
within a  general theory of positive solutions  for parabolic
equations with unbounded coefficients (of particular importance,
if we do not impose any asymptotic fall off restrictions),
\cite{krzyz,kal,bes,aron}.   The fundamental solutions, if their
existence can be  granted, usually live on  space-time strips,
and generally do not admit  unbounded  time intervals.
We shall disregard these issues at the moment, and assume the
existence of fundamental solutions without any reservations.

By exploiting  the rules of functional (Malliavin, variational)
calculus, under an assumption that we deal with
 a diffusion (in fact, Bernstein) process associated with an adjoint
pair (35), (36), it has been shown in Ref. \cite{abc} that \it
if \rm  the forward
conditonal derivatives of the process exist, then
$(D_+\vec{X})(t)= 2\nu {{\nabla \theta }\over \theta } - \vec{A} =
\vec{b}(\vec{x},t)$, (32)   and:
$${(D^2_+\vec{X})(t) = (D_+\vec{X})(t)\times curl\, \vec{A}(\vec{x})
 + \nabla \Omega (\vec{x},t)  + \nu curl\, (curl\,
 \vec{A}(\vec{x}))}\eqno (38)$$
 where $\vec{X}(0)=0$, $\vec{X}(t)=\vec{x}$, $\times $ denotes
 the vector product in $R^3$ and $2\nu c=\Omega $.

 Since $\vec{B}= curl\, \vec{A}=
 \mu _0\vec{H}$, we identify in the above
 the standard Maxwell equation  for $curl \, \vec{H}$ comprising
magnetic  effects of electric  currents in the system:
 $curl \, \vec{B} = \mu _0 [\dot{\vec{D}} + \sigma _0\vec{E}+
 \vec{J}_{ext}]$ where
 $\vec{D}=\epsilon _0\vec{E}$ while  $\vec{J}_{ext}$ represents
 external electric currents. In case of $\vec{E}=\vec{0}$,
 the external currents  only would  be relevant.
 A demand $curl\, curl\, \vec{A}=
 \nabla (\nabla \vec{A}) - \triangle \vec{A}=0$ corresponds to a
 total absence of such currents, and the Coulomb gauge choice
 $\nabla \vec{A} =0$ would leave us with harmonic functions
 $\vec{A}(\vec{x})$.

Consequently, a correct expression for the magnetically implemented
Lorentz force has appeared
on the right-hand-side of the forward acceleration formula (38),
with the forward drift (32) replacing the classical particle
velocity $\dot{\vec{q}}$ of the classical formula (25).

The above discussion implicitly involves quite sophisticated
mathematics, hence it is instructive to see that
we can bypass  the apparent complications  by directly invoking
the universal definitions (7) and (11) of conditional expectation
values, that are based on exploitation of the It\^{o}
formula only. Obviously, under an assumption that
the  Markovian diffusion process   with well defined
transition probability densities $p(\vec{y},s,\vec{x},t)$
and $p_*(\vec{y},s,\vec{x},t)$, does exist.

We shall utilize an  obvious generalization of canonical
definitions (19) of both forward and backward drifts of the
diffusion process defined by the adjoint parabolic
pair (18), as suggested by (32) with $\vec{A}=\vec{A}(\vec{x})$:
$${\vec{b}=2\nu {{\nabla \theta }\over \theta } - \vec{A}
\, \, \, , \, \, \,
\vec{b}_* = - 2\nu {{\nabla \theta _*}\over  {\theta _*}}
- \vec{A} \enspace .}\eqno (39)$$
We also demand that the corresponding adjoint equations (35), (37)
\it are \rm  solved by $\theta $ and $\theta _*$ respectively.

Taking for granted that identities $(D_+\vec{X})(t) = \vec{b}
(\vec{x},t),\, \vec{X}(t)=\vec{x}$ and $(D_-\vec{X})(t)=
\vec{b}_*(\vec{x},t)$ hold true, we can easily evaluate the
forward and backward accelerations (substitute (39), and exploit
the equations (35), (37)):
$${(D_+\vec{b})(\vec{X}(t),t)=\partial _t\vec{b} +
(\vec{b}\nabla )\vec{b} + \nu \triangle \vec{b} =}\eqno (40)$$
$$= \vec{b}\times \vec{B}  + \nu \, curl\, \vec{B}
+ \nabla \, \Omega $$
and
$${(D_-\vec{b}_*)(\vec{X}(t),t)= \partial _t \vec{b}_* +
(\vec{b}_*\nabla
)\vec{b}_* - \nu \triangle \vec{b}_*  =}\eqno (41)$$
$$= \vec{b}_*\times \vec{B} -\nu \, curl\, \vec{B}
+ \nabla \, \Omega \enspace .$$

 Let us notice that the forward and backward
acceleration formulas \it  do not \rm coincide as was the case
before ( cf. Eq. (8), (12)). There is a definite time-asymmetry
in the local description of the diffusion process in the
presence  of general magnetic fields, unless  $curl \, \vec{B} =0$.
  The quantity which is  explicitly
time-reversal invariant can be easily introduced:
$${\vec{v}(\vec{x},t)={1\over 2}(\vec{b} +
\vec{b}_*)(\vec{x},t)\Rightarrow }\eqno (42)$$
$${1\over 2}(D^2_+ + D^2_-)(\vec{X}(t))=
\vec{v}\times \vec{B} + \nabla \, \Omega \enspace .$$
As yet there is no  trace of Lorentzian electric forces, unless
extracted from the term $\nabla \, \Omega (\vec{x},t)$.  This step
we shall accomplish in Section 4.

For a probability density $\theta _* \theta =\rho $ of the
related Markovian diffusion process, \cite{zambr,olk2},
we would have fulfilled
both the  Fokker-Planck and the continuity equations:
$\partial _t\rho  = \nu \triangle \rho -
\nabla (\vec{b}\rho )= - \nabla (\vec{v}\rho )=
-\nu \triangle \rho - \nabla (\vec{b}_*\rho )$, as before
(cf. Section 1).

In the above, the equation (41) can be regarded as the Burgers
equation with a general external magnetic (plus other
external force contributions if necessary) forcing,
and its definition is an
outcome of the underlying  mathematical structure related to
the adjoint pair (33), (37) of parabolic equations.
Our construction shows that the  solution  of the magnetically
forced Burgers equation needs to be sought  in the form  (39).

\section{Schr\"{o}dinger's interpolation in a constant magnetic
field and quantally inspired generalisations}

Presently, we shall confine our attention to the simplest case
of a constant   magnetic field, defined by the vector potential
$\vec{A}=\{ -{B\over 2}x_2, +{B\over 2}x_1, 0  \}$.
Here,  $\vec{B}=
\{ 0, 0, B\}$, $\nabla \vec{A}= 0$ and  $ curl\, \vec{B}= \vec{0}$
which significantly  simplifies formulas (32)-(42).

As emphasized before, most of our discussion was based on the
existence assumption for fundamental solutions of the
(adjoint) parabolic equations (33), (37).
 For magnetic fields, which do
not vanish at spatial infinities (hence for our "simplest"
choice), the situation becomes
rather complicated.  Namely, an expression for
$${c_{\vec{A}}(\vec{x},t)= c(\vec{x},t) - {{B^2}\over {16\nu }}
(x^2_1+x^2_2)}\eqno (43)$$
includes a \it repulsive \rm harmonic oscillator contribution.

For the existence of a well defined Markovian diffusion process
it appears necessary that a nonvanishing contribution from  an
unbounded from above $c(\vec{x},t)$  would counterbalance the
harmonic repulsion.
To see that this \it must be \rm the case, let us
formally constrain $\theta (\vec{x},t)=exp(\Phi (\vec{x},t))$
to yield (in accordance with (9)) the identity:
$${c(\vec{x},t) = \partial _t\Phi  +
\nu  [\nabla \Phi ]^2 +
\nu \triangle \Phi = 0}\eqno (44)$$
Then, we deal with the simplest version of the adjoint system
(35), (37) where, in view of $\nabla \vec{A}=0=c$, there holds:
$${\partial _t\theta =
-\nu [\nabla - {1\over {2\nu }}\vec{A}]^2\theta =
-\nu \triangle \theta + \vec{A}\nabla \theta -
{1\over {4\nu }}[\vec{A}]^2\theta }
\eqno (45)$$
$$\partial _t\theta _*=\nu [\nabla +{1\over {2\nu }}\vec{A}]^2
\theta _*
= \nu \triangle \theta _*   + \vec{A}\nabla \theta _* +
{1\over {4\nu }}[\vec{A}]^2\theta _* \enspace .$$

With  our choice, $curl \, \vec{A}=\{ 0, 0, B\} $, the
equations (45) \it do not  \rm possess a fundamental solution,
which would be well defined for \it all \rm
$(\vec{x},t)\in R^3\times R^+$:
 everything because of the harmonic repulsion term in the
 forward parabolic equation.
 In the Appendix, we shall prove that
 the function $k(\vec{y},s,\vec{x},t)$, :
$${k(\vec{y},s,\vec{x},t)=
[{B\over {4\pi sin({1\over 2}B(t-s))}}]
({1\over {2\pi (t-s)}})^{1/2}}
\eqno (46)$$
$$\times exp\{ - {1\over {2(t-s)}}(x_3-y_3)^2 - {B\over 4}
cot({B\over 2}(t-s)) [(x_2-y_2)^2+ (x_1-y_1)^2] -
 {B\over 2}(x_1y_2-x_2y_1)\} $$
\it only \rm when restricted to
 times $t-s\leq {\pi \over B}$  is an acceptable example of a
  \it unique \rm positive (actually, positivity extends to
times $t-s\leq {{ 2\pi }\over B}$) fundamental solution of
the system (44),  (rescaled to yield
$\nu \rightarrow {1\over 2}$).
Here, formally, (46) can be obtained from the expression (29)
by the  replacement $\vec{A} \rightarrow -i\vec{A}$.

An immediate insight into a harmonic repulsion  obstacle
 can be achieved after an $x-y$-plane rotation
 of Cartesian coordinates:
$x_1'=x_1cos(\omega t) - x_2sin(\omega t), \,
x_2'=x_1sin(\omega t)+x_2cos(\omega t), \, x_3'=x_3,\, t'=t$ with
$\omega ={B\over {4\sqrt{\nu }}}$. Then, equations (45) get
transformed into  an adjoint pair:
$${\partial_{t'} \theta = -\nu \triangle '\theta -
\omega ^2(x_1'^2+x_2'^2)\theta }\eqno (47)$$
$$\partial _{t'}\theta _*=\nu \triangle '\theta _*  +
\omega ^2(x_1'^2+x_2'^2)\theta _*\enspace .$$
Notice that the transformation $\omega \rightarrow i\omega$
would replace  repulsion in (47) by the harmonic attraction.
On the other hand, we can get rid of the repulsive term
by assuming that $c(\vec{x},t)$, (43) does not identically vanish.
For example, we can formally demand that, instead of (44),
$c(\vec{x},t)= +{B^2\over {8\nu }}(x_1^2+x_2^2)$ plays the
r\^{o}le of  an electric potential.
Then,  harmonic attraction  replaces repulsion
 in the final form of equations (35), (37).

As a byproduct,  we are given a  transition probability
density of the diffusion process governed by the adjoint system
(cf. (28)):
$${\partial _t\theta =
-\nu \triangle \theta + \vec{A}\nabla \theta }
\eqno (48)$$
$$\partial _t\theta _*=
 \nu \triangle \theta _*   + \vec{A}\nabla \theta _* \enspace .$$
with $\vec{A}={B\over 2}\{ -x_2, x_1, 0 \}$.
Namely, by means of the previous $x-y$-plane rotation, (48)
is transformed into a pair of time adjoint heat equations:
$${\partial _{t'}\theta = - \nu \triangle '\theta \, \, \, ,\, \, \,
\partial _{t'}\theta _* = \nu \triangle '\theta _* }\eqno (49)$$
whose fundamental solution is the standard heat kernel.

Finding explicit analytic solutions of rather involved
equations (35), (37) is a formidable task on its own,
in contrast to much simpler-unforced or conservatively forced
dynamics issue.

Interestingly, we can produce a number of examples  by invoking
the quantum Schr\"{o}dinger dynamics.   This
quantum inspiration has been proved to be very useful in the past,
\cite{zambr,zambr1}.
At this point, we shall follow the  idea of Ref. \cite{olk2}
where the strategy developed for solving the Schr\"{o}dinger boundary
data problem has been applied to quantally induced stochastic
processes (e.g.  Nelson's diffusions, \cite{nel,nel1}).
They were considered  as  a particular case  of the
general theory appropriate for nonequilibrium statistical
physics processes as governed by the adjoint pair (18), and
exclusively in conjunction with Born's statistical postulate
in quantum theory.

The Schr\"{o}dinger picture  quantum evolution  is then
consistently representable as a Markovian diffusion process. All
that follows from the previously outlined Feynman-Kac kernel route,
\cite{nel,nel1,zambr,olk2,blanch,olk,olk1}, based on exploiting
the adjoint pairs of parabolic equations.
However, the respective semigroup theory has been developed for
pure gradient drift fields, hence without reference to
any impact of electromagnetism on the pertinent diffusion process.
And electromagnetism is definitely ubiquitous in the world of
quantum phenomena.

Let us start from an ordinary Schr\"{o}dinger equation
 for a charged particle in  an arbitrary
external electromagnetic field, in its standard dimensional form.
To conform with the previous notation let us absorb the charge
$e$ and mass $m$ parameters in the definition of $\vec{A}(\vec{x})$
and the potential $\phi (\vec{x})$, e.g. we consider
$B$ instead of ${e\over m}B$ and $\phi $ instead $\phi /m$.
Additionally, we set $\nu $ instead of ${\hbar \over {2m}}$.
Then, we have:
$${i\partial _t \psi (\vec{x},t)= - \nu (\nabla  -
{i\over {2\nu }}\vec{A})^2 \psi (\vec{x},t) + {1\over {2\nu }}
\phi (\vec{x})\psi (\vec{x},t)\enspace . }\eqno (50)$$

The standard Madelung substitution   $\psi = exp(R+iS)$ allows
to introduce the real functions $\theta = exp(R+S)$ and
$\theta _*=exp(R-S)$ instead of complex ones
$\psi, \overline {\psi }$.
They are solutions of an adjoint  parabolic system (35), (37),
where the impact of (50) is encoded in a specific functional form
of, otherwise arbitrary potential $c(\vec{x},t)$:
$${c(\vec{x},t)= {1\over {2\nu }} \Omega (\vec{x},t)=
{1\over {2\nu }} [2Q(\vec{x},t)-\phi (\vec{x})]}
\eqno (51)$$
$$Q(\vec{x},t)=2\nu ^2{{\triangle \rho ^{1/2}(\vec{x},t)}\over {\rho
^{1/2}(\vec{x},t)}}=2\nu ^2 \bigl [ \triangle R(\vec{x},t) + [\nabla
R(\vec{x},t)]^2 \bigr ]\enspace . $$

The quantum probability density $\rho (\vec{x},t)=
\psi (\vec{x},t)\overline {\psi }(\vec{x},t)=
\theta (\vec{x},t)\theta _*(\vec{x},t)$  displays a factorisation
 $\rho =\theta \theta _*$ in terms of solutions of adjoint parabolic
 equations, which we recognize to be characteristic for probabilistic
 solutions  (Markov diffusion processes)
 of the Schr\"{o}dinger boundary data problem (cf. Section 1),
 \cite{zambr,blanch,olk2,olk}.
It is  easy to verify the validity of the Fokker-Planck equation
whose forward drift has the  form (38).  Also, equations (40), (41)
do follow with $\Omega =2Q - \phi $.

By defining $\vec{E}=-\nabla \phi $
(with $\phi $ utilised instead of
${e\over m}\phi $), we immediately arrive at the complete Lorentz
force contribution in all acceleration formulas
(before, we have used $curl \, \vec{B}=0$):
$${\partial _t\vec{b} + (\vec{b}\nabla )\vec{b} +
\nu \triangle \vec{b} = \vec{b}\times \vec{B} + \vec{E} +
\nu curl \, \vec{B} +  2\nabla Q
}\eqno (52)$$
$$\partial _t\vec{b}_* + (\vec{b}_*\nabla )\vec{b}_* -\nu \triangle
\vec{b}_* = \vec{b}_* \times \vec{B} + \vec{E} - \nu curl \, \vec{B}
+ 2\nabla Q$$
Moreover, the velocity field named the current velocity of the
flow, $\vec{v}={1\over 2}(\vec{b} + \vec{b}_*)$,
enters  the familiar local conservation laws
(see also \cite{blanch} for a discussion of how the "quantum potential"
$Q$ affects such  laws in case of the standard Brownian
motion and Smoluchowski-type diffusion processes)
$${\partial \rho = - \nabla (\vec{v}\rho )}\eqno (53)$$
$$\partial _t\vec{v} +(\vec{v}\nabla )\vec{v} =
\vec{v}\times \vec{B} + \vec{E}  + \nabla Q \enspace .$$

A comparison with (33)-(43) shows that equations (50)-(53) can
be regarded  as  the  specialized version  of the general external
forcing problem with an explicit electromagnetic (Lorentz
force-inducing) contribution and an
arbitrary  term  of non-electromagnetic origin, which we denote by
$c(\vec{x},t)$ again. Obviously,  $c$ is represented in (51),
 by ${1\over {\nu }}Q(\vec{x},t)$.

We have therefore arrived at the
following ultimate generalization of the adjoint parabolic
system (18), that encompasses the nonequilibrium statistical
physics and essentially quantum evolutions on an equal footing
(with no clear-cut discrimination between these options, as in
Ref. \cite{olk2})
and gives rise to an external (Lorentz)  electromagnetic forcing:
$${\partial _t\theta (\vec{x},t)=[ - \nu (\nabla  -
{1\over {2\nu }}\vec{A})^2 -
{1\over {2\nu }}\phi (\vec{x}) + c(\vec{x},t)] \theta
(\vec{x},t)}\eqno (54)$$
$$\partial _t\theta _*(\vec{x},t)= [\nu (\nabla +
 {1\over {2\nu }}\vec{A})^2
 + {1\over {2\nu }}\phi (\vec{x}) -
c(\vec{x},t)]\theta _*(\vec{x},t)\enspace . $$
A subsequent  generalisation encompassing time-dependent
electromagnetic  fields is immediate.

The adjoint parabolic pair  (54) of equations can  thus be
regarded to determine a Markovian diffusion
process in exactly the same way as (18) did.
If only a suitable choice of  vector and scalar potentials in (54)
guarrantees  a continuity and positivity of the involved semigroup
kernel (take the Radon-Nikodym density of the form (34), with
$\Omega  \rightarrow -\phi + \Omega $ , and integrate with
respect to the  conditional Wiener measure), then the mere
knowledge of such integral  kernel
suffices  for the implementation of steps (18)-(22), with
$u\rightarrow \theta _*$, $v\rightarrow \theta $.  To this end
it is not at all necessary that $k(\vec{x},s,\vec{y},t)$ is a
fundamental solution  of (54). A sufficient condition  is that
the semigroup kernel is a
continuous (and positive) function. The kernel  may not even be
differentiable, see e.g. Ref. \cite{olk2} for a discussion of
that issue  which is typical for quantal situations.

After adopting (54) as the principal dynamical ingredient of
the electromagnetically forced Schr\"{o}dinger interpolation,
we  must slightly adjust  the emerging acceleration
formulas. Namely, they have the form (52) , but we
need to  replace $2Q(\vec{x},t)$  by, from now on
 arbitrary,  potential $\Omega (\vec{x},t)= 2\nu c(\vec{x},t)$.
The  second equation in (53) also takes a new  form:
$${\partial _t\vec{v} + (\vec{v}\nabla )\vec{v} =
\vec{v} \times \vec{B} + \vec{E} + \nabla (\Omega  - Q)}
\eqno (55)$$
see e.g. Ref. \cite{blanch} for more detailed explanation of
this step.  The presence in (54) of the  density-dependent
$- \nabla Q$ term  finds its origin  in the  identity
$\vec{b}-\vec{b}_* = 2\nu \nabla \rho (\vec{x},t)$ and is a necessary
consequence of the involved (forced in the present case)  Brownian
motion, see e.g. \cite{gei,garb1,vig}.

Finally, the second equation  (52) with $\Omega $ replacing $2Q$
is the most general form of the Burgers equation with  an
external forcing, where  the electromagnetic (Lorentz force)
contribution has been extracted for convenience. Solutions of
this equation must be sought for  in the form (39), which
 generalizes the logarithmic Hopf-Cole transformation to
 non-gradient drift  fields.
 Equations (54) are the associated
 parabolic partial differential (generalised heat) equations, that
 completely determine probabilistic solutions (Markovian diffusion
 processes) of the  Schr\"{o}dinger boundary data (interpolation)
 problem, for which the forced Burgers velocity fields play the
 r\^{o}le of  backward drifts of the process.

\section*{Appendix}

General criterions, allowing to decide when the semigroup kernel
associated  with the adjoint pair (54) is a continuous and
positive function, belong to a specialised branch of mathematical
theory  of parabolic partial differential equations and
mathematical physics. A list of restrictions suitable for our
case can be found in Section 4 of Ref. \cite{abc}.

To give a flavour of a typical argumentation, we shall
present  a more selective discussion of conditions under
which the  quantally inspired parabolic problem (33), (37), (51)
admits fundamental solutions. Let us point out that quite
generally  fundamental solutions do not exist (the kernels do !)
in such case, cf. \cite{olk2,olk1}.

Let us rescale the problem (50) so that we are interested in:
$${i\partial _t \psi (\vec{x},t)= - {1\over 2} (\nabla  -
i\vec{A})^2 \psi (\vec{x},t) +
\phi (\vec{x})\psi (\vec{x},t)\enspace . }\eqno (A.1)$$
with the initial data $\psi_0(\vec{x}) \in L^2(R^3)$. We assume
that $A_i, i=1,2,3, \in C^1(R^3)$ i.e. have continuous first
derivatives, while $\phi (\vec{x}) \in C(R^3)$ and is bounded
from below, e.g. $\phi (\vec{x}) > - M$ for some positive $M$.
It is well known that under these restrictions the operator
$H={1\over 2}(i\nabla + \vec{A})^2 +\phi $  is essentially
selfadjoint  on $C^{\infty }_c(R^3)$,  i.e. the space of smooth
functions with a compact support, \cite{reed}. Thus
$\psi _t=exp(-it\overline {H})\psi _0$ is a solution of (A.1)
for every time  $t$ and any initial condition $\psi _0\in
D(\overline {H})$. Here, $\overline {H}$ denotes the closure of an
operator $H$.

Let us fix  some $T>0$ and let us assume that $\psi (\vec{x},t)\neq 0$
for all  $\vec{x}\in R^3$ and $t\in [0,T]$. Then, the Madelung
substitution $\psi =exp(R+iS)$ implies:
$${\partial _t R=-\nabla R\nabla S - {1\over 2} \triangle S +
{1\over 2} \nabla\vec{A} + \vec{A}\nabla R}\eqno (A.2)$$
$$\partial _t S= {1\over 2} (\nabla R)^2 - {1\over 2} (\nabla S)^2 +
{1\over 2} \triangle R  + \vec{A} \nabla S - {1\over 2} \vec{A}^2
- \phi \enspace .$$
At the same time, the real-valued functions $\theta =exp(R+S)$ and
$\theta _*=exp(R-S)$ obey the adjoint parabolic equations (to be
compared with (35), (37)):
$${\partial _t\theta _*= {1\over 2} \triangle \theta _* +
\vec{A}\nabla \theta _* - c_{\vec{A},\phi }\theta _* }\eqno (A.3)$$
$$\partial _t \theta = - {1\over 2} \triangle \theta +
\nabla (\vec{A} \theta ) + c_{\vec{A},\phi }\theta $$
where
$${c_{\vec{A},\phi } = {1\over 2}\nabla \vec{A} - {1\over 2}
\vec{A}^2 + 2Q-\phi }\eqno (A.4)$$
$$2Q={{\triangle \rho ^{1/2}}\over {\rho ^{1/2}}}= (\nabla R)^2 +
\triangle R \enspace .  $$

Let us furthermore assume that $\partial A_i/\partial x_j$ and
$c_{\vec{A},\phi }$ are locally H\"{o}lder continuous  and
moreover that
$${|A_i| \leq K_1(\vec{x}^2 + 1)^{1/2}}\eqno (A.5)$$
$$|c_{\vec{A},\phi }| < K_2(\vec{x}^2 +1)$$
for all $i=1,2,3$. Then, for a certain $T_0$ which may be smaller
that the previously chosen $T$, there  exists   a unique
fundamental solution $k(\vec{y},s,\vec{x},t)$  defined for all
$\vec{x}, \vec{y} \in R^3$,  \it but \rm confined to a time interval
$0\leq s< t\leq T_0$.\\

The solution has  the following properties:\\
(i) For any $\eta \in C_c(R^3)$, $\theta _*(\vec{x},t) =\int_{R^3}
k(\vec{y},s,\vec{x},t)\eta (\vec{y})d^3y$ is a regular solution of
the   forward equation (A.3) such that $lim_{t\downarrow s}
\theta _*(\vec{x},t) = \phi (\vec{x})$.\\
(ii) $k$ is a regular solutions of the forward equation (A.3) in
$(\vec{x},t)$ variables and a regular solution of the backward
equation (A.3) in the $(\vec{y},s)$ variables. \\
(iii) $\int_{R^3} k(\vec{y},s,\vec{x},t)d^3y \leq C_1exp(C_2x^2)$
for some positive $C_1$ and $C_2$.\\
(iv) $k$ is strictly positive i.e. $k>0$.\\
(v) $k$ obeys the Chapman-Kolmogorov equation
$\int_{R^3} k(\vec{y},s,\vec{z},u)k(\vec{z},u,\vec{x},t)d^3z=
k(\vec{y},s,\vec{x},t)$ for all $s<u<t\leq T_0$.\\

In the above, properties (i) and the first part of (ii) are
in fact a definition of a fundamental solution, \cite{fried}.
The second property (ii)  and the estimate (iii) are canonical
properties of the fundamental solution,\cite{kal,aron}.
The property (iv)  comes as a consequence of another estimate
(Lemma 2 in Ref. \cite{aron}): for every $\epsilon \in (0,t)$
and every $x\in R^3$ there exist constants $\Lambda $ and $\mu $
such that   $k(\vec{y},s,\vec{x},t) \geq \Lambda exp[-\mu
(\vec{x}-\vec{y})^2]$ for all $\vec{y} \in R^3 $ and all
$s\in [0,t-\epsilon ]$. The Chapman-Kolmogorov equation
(property (v)) follows from the uniqueness of positive solutions
of the forward equation (A.3), \cite{aron,kal}.  \\
Consequently, what needs to be proven is the \it uniqueness \rm of
the integral kernel $k$. \\
To this end, let us  assume that there are two fundamental solutions.
Then, for any $\eta \in C_c(R^3)$ we have
$${u_i(\vec{x},t)=\int_{R^3} k_i(\vec{y},s,\vec{x},t)\eta
(\vec{y})d^3y}\eqno  (A.6)$$
for $i=1,2$, and $u_i(\vec{x},t)$ are the regular solutions such that
$u_(\vec{x},0)=\eta (\vec{x})$. \\
Let us define $u=u_2-u_1$. Then, for $k_0>C_2$ there holds:
$${\int_0^{T_ 0} \int_{R^3} |u(\vec{x},t)|exp(-k_0\vec{x}^2) d^3xdt
\leq }\eqno (A.7)$$
$$\leq \int_0^{T_0} \int_{R^3} \int_{R^3} (k_1 + k_2)
(\vec{y},s,\vec{x},t)
\eta (\vec{y}) exp(-k_0\vec{x}^2) d^3y d^3x dt\enspace .$$
But we have $\int_{R^3} k_i(\vec{y},s,\vec{x},t) d^3y \leq
C_1 exp(C_2\vec{x}^2)$ for $i=1,2$. Consequently:
$${\int_0^{T_0} \int_{R^3} |u(\vec{x},t)| exp(-k_0x^2) d^3x dt\leq
2||\eta ||_{sup}\,  C_1 T_0 \int_{R^3} exp[-(k_0-C_2)\vec{x}^2] d^3x
<\infty \enspace .}\eqno (A.8)$$
By Theorem AB2 of Ref. \cite{aron}, in view of  $u(\vec{x},0)=0$ we
here arrive at $u(\vec{x},t)\equiv 0$.
Since $\eta $ was arbitrary, we conclude
that $k_1\equiv k_2$. The uniqueness has been proved.

One may argue about the seemingly spurious time interval restriction
to $T_0<T$. This time domain limitation for the  kernel $k$ can be
easily justified. Namely, let us consider the forward equation
(cf. A.3)  $\partial _tu={1\over 2}\triangle u +
\vec{A}\nabla u + cu $
 and define $v=uexp[-g(\vec{x},t)]$ where $g(\vec{x},t) =
{\vec{x}^2\over {1-at}}$ and $a>0$. The newly introduced function $v$
satisfies:
$${\partial _tv={1\over 2}\triangle v + (\vec{B}\nabla )v+ c_0v}
\eqno (A.9)$$
with $\vec{B} =\vec{A} + {1\over 2}\nabla g$ and $c_0
= c_{\vec{A},\phi } + {1\over 2} (\nabla g)^2 + {1\over 2}
\triangle g + (\vec{A}\nabla )g - \partial _tg$. Because of
$\partial _tg= {{a\vec{x}^2}\over {(1-at)^2}}$, by taking $a$
large enough we
can make the equation (A.9) "dissipative", \cite{eidel}, and so
construct its fundamental solution $k_0$. A fundamental solution of
the forward equation (A.3)  is the given by:
$${k(\vec{y},s,\vec{x},t)= k_0(\vec{y},s,\vec{x},t) exp[
g(\vec{x},t)-
g(\vec{y},s)] \enspace . } \eqno (A.10)$$
However if $a$ is large, then $t$ can run only a small interval
$[0,1/a]$.

All the above features of the fundamental solution, (i)-(v),
 can be explicitly checked (by inspection) for the kernel (46),
 associated with the parabolic pair (45) in a constant magnetic field.
 They hold true for times $0\leq s<t\leq T_0< \pi $, except for the
 property  (iii) which is true \it only \rm if $T<{\pi \over 2}$.
 Indeed:
 $${{1\over {2\pi sin(t-s)}} \int_{R^2} exp\{ -{{cot(t-s)}\over 2}
 [(x_1-y_1)^2 + (x_2-y_2)^2]\}  \, exp (-x_2y_1+x_1y_2) dy_1dy_2}
 \eqno (A.11)$$
 $$= cos(t-s)\, exp[{{tan(t-s)}\over 2}(x_1^2+x_2^2)]\enspace .$$
 Since $[2\pi (t-s)]^{-1/2}\int_R exp[-(x_3-y_3)^2/2(t-s)]dy_3 = 1$,
 to achieve (iii) it suffices to put  $C_1=1$ and
 $C_2={1\over 2}tan T_0$, provided $T_0<{\pi \over 2}$.

 The uniqueness of the kernel  for time $T_0<{\pi \over 2}$ is proved
 along the same lines as this for the fundamental solution of (A.3).

\vskip0.5cm
{\bf Acknowledgement}: Two of the  authors (P. G. and R. O. )
received  a financial support from the KBN research grant
No 2 P302 057 07.  P. G. would like to express his gratitude to
Professors Ana Bela Cruzeiro and Jean-Claude Zambrini for
enlightening discussions.

\end{document}